\documentclass[aps,prd,showpacs,nofootinbib,preprintnumbers]{revtex4}
\usepackage{amsmath}
\usepackage{amssymb}
\usepackage{amsfonts}
\usepackage{graphicx}
\usepackage{bm}

\newcommand{\be}{\begin{equation}}
\newcommand{\ee}{\end{equation}}
\newcommand{\bea}{\begin{eqnarray}}
\newcommand{\eea}{\end{eqnarray}}
\newcommand{\beaa}{\begin{eqnarray*}}
\newcommand{\eeaa}{\end{eqnarray*}}

\newcommand{\nn}{\nonumber \\}

\begin{document}

\tolerance=5000

\preprint{YITP-10-91}

\title{Screening of cosmological constant in non-local gravity}
\author{Shin'ichi Nojiri$^{1,2}$, Sergei D. Odintsov$^3$, Misao
Sasaki$^{4,5}$
 and Ying-li Zhang$^4$}

\affiliation{
~\\
${}^1$Department of Physics, Nagoya University,~Nagoya 464-8602, Japan
~\\
\\
${}^2$Kobayashi-Maskawa Institute for the Origin of Particles and the
Universe,
Nagoya University, Nagoya 464-8602, Japan
~\\
\\
${}^3$Institucio Catalana de Recerca i Estudis Avancats (ICREA)
and Institut de Ciencies de l'Espai (IEEC-CSIC),
Campus UAB, Facultat de Ciencies, Torre C5-Par-2a pl, E-08193 Bellaterra
(Barcelona), Spain
~\\
\\
${}^4$Yukawa Institute for Theoretical Physics, Kyoto University,
Kyoto 606-8502, Japan
~\\
\\
${}^5$Korea Institute for Advanced Study,
Seoul 130-722, Republic of Korea
~\\
}

\date{\today}
\begin{abstract}
We discuss a possible mechanism to screen a cosmological constant in
non-local gravity. We find that in a simple model of of non-local
gravity with the Lagrangian of the form, $R+f(\Box^{-1}R)-2\Lambda$
where $f(X)$ is a quadratic function of $X$, there is a flat
spacetime solution despite the presence of the cosmological constant
$\Lambda$. Unfortunately, however, we also find that this solution
contains a ghost in general. Then we discuss the condition to avoid
a ghost and find that one can avoid it only for a finite range of
`time'. Nevertheless our result suggests the possibility of solving
the cosmological constant problem in the context of non-local
gravity.
\end{abstract}

\maketitle
\section{Introduction}

The study of string/M-theory or conventional quantum gravity, for instance,
using the effective action approach often leads to the appearance of
non-local gravitational terms. These terms are not easy to deal with,
nevertheless,
they maybe important in the applications for early-time and/or late-time
universe. In particular, they maybe responsible for early/late-time
acceleration. Recently,
 there has been increasing interest in various aspects of
non-local gravity\cite{Nojiri:2007uq,Jhingan:2008ym,Capozziello:2008gu,
Cognola:2009jx,Deser:2007jk,nonlocal}.
Various different theories of non-local gravity
have been proposed in these works and their
accelerating FLRW cosmology was investigated. It was found that in the same
way as other modified gravity theories
non-local gravity may (at least, qualitatively)
lead to a unified description of the early-time inflation
and the present dark energy era
(for review of such an approach, see \cite{review}).

The present paper is devoted to a related aspect of non-local gravity:
the possible solution to the cosmological constant problem.
Indeed, it was suggested sometime ago that
the cosmological constant problem may be solved by
non-local gravity~\cite{ArkaniHamed:2002fu}.
 Here we give an explicit example
for such a proposal. In the next section the model of non-local gravity is
considered and its scalar-tensor presentation is given. Its ghost-free
condition is analyzed in section III. A simple example which
is ghost-free and which provides the solution to the cosmological constant
problem is given in section IV. Summary is given in Discussion section.
In Appendix we propose a new model of non-local gravity with
a Lagrange constraint multiplier.

\section{non-local gravity}
The starting action of the non-local gravity is given by
\be
\label{nl1}
S=\int d^4 x \sqrt{-g}\left\{
\frac{1}{2\kappa^2}\left\{ R\left(1 + f(\Box^{-1}R )\right) -2 \Lambda
\right\}
+ \mathcal{L}_\mathrm{matter} \left(Q; g\right)
\right\}\, .
\ee
Here $f$ is some function, $\Box$ is the d'Alembertian for scalar field,
 $\Lambda$
is a cosmological constant
and $Q$ stands for the matter fields.
The above action can be rewritten by introducing two scalar fields $\psi$
and $\xi$ in the following form,
\bea
\label{nl2}
S&=&\int d^4 x \sqrt{-g}\left[
\frac{1}{2\kappa^2}\left\{R\left(1 + f(\psi)\right)
- \xi\left(\Box\psi - R\right) - 2 \Lambda \right\}
+ \mathcal{L}_\mathrm{matter}  \right] \nn
&=&\int d^4 x \sqrt{-g}\left[
\frac{1}{2\kappa^2}\left\{R\left(1 + f(\psi)+\xi\right)
 + g^{\mu\nu}\partial_\mu \xi \partial_\nu \psi  - 2 \Lambda \right\}
+ \mathcal{L}_\mathrm{matter}
\right]
\, .
\eea
By varying the above with respect to $\xi$, we obtain
$\Box\psi=R$ or $\psi=\Box^{-1}R$.
Substituting the above equation into (\ref{nl2}), one re-obtains
(\ref{nl1}).

By varying (\ref{nl2}) with respect to the metric $g_{\mu\nu}$ gives
\bea
\label{nl4}
0 &=& \frac{1}{2}g_{\mu\nu} \left\{R\left(1 + f(\psi) + \xi\right)
 + g^{\alpha\beta}\partial_\alpha \xi \partial_\beta \psi - 2 \Lambda
\right\}
 - R_{\mu\nu}\left(1 + f(\psi) +\xi\right) \nn
&& - \frac{1}{2}\left(\partial_\mu \xi \partial_\nu \psi
+ \partial_\mu \psi \partial_\nu \xi \right)
 -\left(g_{\mu\nu}\Box - \nabla_\mu \nabla_\nu\right)\left( f(\psi) +
\xi\right)
+ \kappa^2T_{\mu\nu}\, .
\eea
On the other hand, the variation with respect to $\psi$ gives
\be
\label{nl5}
0=\Box\xi- f'(\psi) R\, .
\ee
Now we assume the FLRW metric
\be
\label{nl6}
ds^2 = - dt^2 + a(t)^2 \delta_{ij}dx^idx^j\, ,
\ee
and the scalar fields $\psi$ and $\xi$ only depend on time.
Then Eq.~(\ref{nl4}) takes the following form,

\bea \label{einstein1} 0 &=& - 3 H^2\left(1 + f(\psi) + \xi\right) -
\frac{1}{2}\dot\xi \dot\psi
 - 3H\left(f'(\psi)\dot\psi + \dot\xi\right) + \Lambda
+ \kappa^2 \rho\, ,\\
\label{einstein2} 0 &=& \left(2\dot H + 3H^2\right) \left(1 +
f(\psi) + \xi\right) - \frac{1}{2}\dot\xi \dot\psi +
\left(\frac{d^2}{dt^2} + 2H \frac{d}{dt} \right) \left( f(\psi) +\xi
\right)
 - \Lambda + \kappa^2 P\,,
\eea
where $H=\dot a/a$, and $\rho=-T^0{}_0$ and $P=T^i{}_i/3$ are
the energy density and pressure, respectively,
of the matter fields.
On the other hand, the scalar equations are
\bea
\label{psieq}
0 &=& \ddot \psi + 3H \dot \psi + 6 \dot H + 12 H^2 \, , \\
\label{xieq} 0 &=& \ddot \xi + 3H \dot \xi + \left( 6 \dot H + 12
H^2\right)f'(\psi) \, . \eea

\section{ghost-free condition}
Let $\tilde g_{\mu\nu}$ be the original metric discussed in the
previous section. We make a conformal transformation to the Einstein
frame,
\begin{eqnarray}
\tilde g_{\mu\nu}=\Omega^2g_{\mu\nu}\,,
\quad
\tilde R=\frac{1}{\Omega^{2}}\left[R-6\left(\Box\ln\Omega
+g^{\mu\nu}\nabla_\mu\ln\Omega\nabla_\nu\ln\Omega\right)\right]\,,
\end{eqnarray}
with
\begin{eqnarray}
\Omega^2=\frac{1}{1+f(\psi)+\xi}\,.
\end{eqnarray}
This gives
\begin{eqnarray}
S&=&\int d^4 x \sqrt{-g}\biggl\{
\frac{1}{2\kappa^2}\left[R-6\left(\Box\ln\Omega
+g^{\mu\nu}\nabla_\mu\ln\Omega\nabla_\nu\ln\Omega\right)
 +\Omega^2 g^{\mu\nu}\nabla_\mu \xi \nabla_\nu \psi-2\Omega^4\Lambda \right]
\cr
&&\qquad
+ \Omega^4\mathcal{L}_\mathrm{matter} (Q;\Omega^2 g)
\biggr\}\, .
\label{efaction0}
\end{eqnarray}
The $\Box\ln\Omega$ term may be discarded because it is a total divergence.
Hence we obtain
\begin{eqnarray}
S&=&\int d^4 x \sqrt{-g}\left\{
\frac{1}{2\kappa^2}
\left[R-6g^{\mu\nu}\nabla_\mu\ln\Omega\nabla_\nu\ln\Omega
 +\Omega^2 g^{\mu\nu}\nabla_\mu \xi \nabla_\nu \psi-2\Omega^4\Lambda \right]
 + \Omega^4\mathcal{L}_\mathrm{matter}  (Q;\Omega^2 g)
\right\}
\cr
&=&
\int d^4 x \sqrt{-g}\left\{
\frac{1}{2\kappa^2}
\left[R-6g^{\mu\nu}\nabla_\mu\phi\nabla_\nu\phi
 +e^{2\phi}g^{\mu\nu}\nabla_\mu \xi \nabla_\nu \psi-2e^{4\phi}\Lambda
\right]
 + e^{4\phi}\mathcal{L}_\mathrm{matter}  (Q;e^{2\phi}g)
\right\}
\, .
\label{efaction}
\end{eqnarray}
where
\begin{eqnarray}
\phi\equiv\ln\Omega=-\frac{1}{2}\ln(1+f(\psi)+\xi)\,.
\end{eqnarray}
Here we note the condition for the gravity to have the normal sign,
\begin{eqnarray}
1+f(\psi)+\xi>0\,.
\label{normalgravity}
\end{eqnarray}

Instead of $\psi$ and $\xi$,
one may regard $\phi$ and $\psi$ to be independent fields.
Then inserting
\begin{eqnarray}
\xi=e^{-2\phi}-(1+f(\psi))
\end{eqnarray}
into (\ref{efaction}), we finally arrive at
\begin{eqnarray}
S&=&\int d^4 x \sqrt{-g}\left\{
\frac{1}{2\kappa^2}
\left[R-6\nabla^\mu\phi\nabla_\mu\phi
 -2\nabla^\mu\phi\nabla_\mu\psi
 -e^{2\phi}f'(\psi)\nabla^\mu \psi \nabla_\mu \psi -2e^{4\phi}\Lambda\right]
 + e^{4\phi}\mathcal{L}_\mathrm{matter}  (Q;e^{2\phi}g)
\right\}
\, .
\label{finalaction}
\end{eqnarray}
In order to avoid a ghost, the determinant of the kinetic term
must be positive. This means
\begin{eqnarray}
\det \left| \begin{array}{cc}
6\quad&\ 1 \\
1\quad&\ e^{2\phi}f'(\psi)
\end{array} \right|=6e^{2\phi}f'(\psi)-1>0\,.
\label{noghost}
\end{eqnarray}
We assume this condition is satisfied. In particular,
$f'(\psi)>0$ is a necessary condition.

To summarize, in terms of the original scalar fields, the condition
for the theory to be healthy is
\begin{eqnarray}
f'(\psi)>\frac{1+f(\psi)+\xi}{6}>0\,.
\label{healthycond}
\end{eqnarray}

\section{simple model}
We consider a simple model for $f(\psi)$:
\begin{eqnarray}\label{psi}
f(\psi)=f_0\psi+f_1\psi^2\,,
\end{eqnarray}
and look for the flat spacetime solution $H=0$. For simplicity we assume
the matter to be absent, $\rho=P=0$.
Then from the scalar equations (\ref{psieq}) and (\ref{xieq}),
the solutions for $\psi(t)$ and $\xi(t)$ can be obtained as
\begin{eqnarray}
\psi(t)=\psi_0+\psi_1t,
\end{eqnarray}
\begin{eqnarray}
\xi(t)=\xi_0+\xi_1t,
\end{eqnarray}
where $\psi_0, \psi_1, \xi_0$ and $\xi_1$ are integral constants.
Without loss of generality we may put $\psi_0=\xi_0=0$.
Also we may assume $\psi_1>0$ because of the time reversal invariance.
Inserting these to Einstein equations (\ref{einstein1}) and
(\ref{einstein2}),
we find
\begin{eqnarray}\label{ein1}
\xi_1=2\Lambda\sqrt{\frac{f_1}{\Lambda}},
\end{eqnarray}
\begin{eqnarray}\label{ein2}
\psi_1=\sqrt{\frac{\Lambda}{f_1}}\,.
\end{eqnarray}
Here we note that $f_1$ must have the same sign as that of $\Lambda$.

Thus we have found that there is a flat spacetime solution in this non-local
gravity
model even under the presence of a cosmological constant. In other words,
the
non-local gravitational effect has successfully shielded the effect of a
cosmological constant.

Now we discuss the condition to avoid the appearance of a ghost in the above
solution. Let us recapitulate the condition (\ref{healthycond}),
\begin{eqnarray}
f'(\psi)>\frac{1+f(\psi)+\xi}{6}>0\,.
\end{eqnarray}
In our model, using Eqs.~(\ref{ein1}) and (\ref{ein2}), the above
condition gives
\begin{eqnarray}
G(t)&\equiv&
t^2\Lambda+t\sqrt{\frac{\Lambda}{f_1}}(f_0-10f_1)+1-6f_0<0,
\label{con2}\\
K(t)&\equiv&t^2\Lambda+t\sqrt{\frac{\Lambda}{f_1}}(f_0+2f_1)+1>0.
\label{con3}
\end{eqnarray}
Below we discuss these conditions in the cases of positive
and negative cosmological constants separately.

\subsection{$\Lambda>0$ case}

First, we discuss the case when $\Lambda>0$. We assume $f_1>0$ for
consistency.
Then the necessary condition to
satisfy (\ref{con2}) is
\begin{equation}
\Delta_G\equiv\Lambda\frac{(f_0+2f_1)^2-4f_1+96f_1^2}{f_1}>0,
\label{DeltaG}
\end{equation}
which directly leads to the condition,
\begin{equation}\label{ghost1}
f_0^2+4f_1(f_0-1)>-100f_1^2.
\end{equation}
Let $t_1$ and $t_2$ be two solutions for $G(t)=0$, we have
\begin{eqnarray}
t_{1,2}= \frac{1}{\Lambda}\sqrt{\frac{\Lambda}{
f_1}}\left\{-\frac{f_0}{2}+5f_1\pm\left[
\frac{f_0^2}{4}+f_1\left(25f_1+f_0-1\right)\right]^{\frac{1}{2}}\right\},
\label{G=0}
\end{eqnarray}
where $t_1<t_2$ is assumed so that the range of $t$ is
between $t_1$ and $t_2$, i.e. $t\in(t_1, t_2)$.

The sufficient condition to satisfy (\ref{con3}) is
\begin{eqnarray}
\Delta_K\equiv\Lambda\frac{(f_0+2f_1)^2-4f_1}{f_1}<0\,,
\label{DeltaK}
\end{eqnarray}
that is,
\begin{eqnarray}\label{ghost3}
f_0^2+4f_1(f_0-1)<-4f_1^2.
\end{eqnarray}

Thus, combining Eqs.~(\ref{ghost1}) and (\ref{ghost3})
together, we conclude that for a positive $\Lambda$ the sufficient
condition to prevent a ghost is
\begin{eqnarray}
-25f_1^2<\frac{f_0^2}{4}+f_1f_0-f_1<-f_1^2\,,
\end{eqnarray}
or
\begin{eqnarray}\label{ghostsum}
\frac{1}{f_1}-25<\frac{f_0^2}{4f_1^2}+\frac{f_0}{f_1}<\frac{1}{f_1}-1\,.
\end{eqnarray}
The range of time $\Delta t_+$ is
\begin{eqnarray}\label{range1}
\Delta t_+= 2\sqrt{\frac{f_1}{\Lambda}}
\left(\frac{f_0^2}{4f_1^2}+\frac{f_0}{f_1}+25-\frac{1}{f_1}\right)^{\frac{1
}{2}}.
\end{eqnarray}

\subsection{$\Lambda<0$ case}
Now we turn to discuss the situation when $\Lambda<0$. In this case
we assume $f_1<0$ for consistency.
The main difference from the case of a positive $\Lambda$ is that
the condition (\ref{DeltaG}) is replaced by $\Delta_K>0$,
\begin{eqnarray}
\Delta_K=\Lambda\left[\frac{(f_0+2f_1)^2}{f_1}-4\right]>0\,,
\end{eqnarray}
which leads to
\begin{eqnarray}
f_0^2+4f_1(f_0-1)>-4f_1^2.
\end{eqnarray}

Since $\Delta_G=\Delta_K+96f_1\Lambda>\Delta_K$, $\Delta_G$ is
always positive if $\Delta_K>0$. Thus $G(t)=0$ has two real
solutions if $K(t)=0$ has two real solutions.

Let $t_3$ and $t_4$ be the solutions for $K(t)=0$, then we have
\begin{eqnarray}
t_{3,4}= \frac{1}{\Lambda}\sqrt{\frac{\Lambda}{
f_1}}\left\{-\frac{f_0}{2}-f_1\pm\left[
\frac{f_0^2}{4}+f_1\left(f_1+f_0-1\right)\right]^{\frac{1}{2}}\right\},
\end{eqnarray}
where we assume $t_3<t_4$. While the solutions for $G(t)=0$
have been already given by $t_{1,2}$ in (\ref{G=0}).

Thus, if we have $t_4<t_1$ or $t_2<t_3$ for a non-zero range of $t$,
the range of $t$ is given by $t\in(t_3, t_4)$.
Both of these requirements yield the same condition,
\begin{eqnarray}
f_0^2+4f_1(f_0-1)<0.
\end{eqnarray}

So we conclude that for $\Lambda<0$, the sufficient condition to
avoid a ghost is
\begin{eqnarray}
-4f_1^2<f_0^2+4f_1(f_0-1)<0,
\end{eqnarray}
or
\begin{eqnarray}\label{nogh}
\frac{1}{f_1}-1<\frac{f_0^2}{4f_1^2}+\frac{f_0}{f_1}<\frac{1}{f_1}\,.
\end{eqnarray}

The range of time $\Delta t_-$ is
\begin{eqnarray}\label{tnegative}
\Delta t_-=2\sqrt{\frac{f_1}{\Lambda}}
\left(\frac{f_0^2}{4f_1^2}+\frac{f_0}{f_1}+1-\frac{1}{f_1}\right)^{\frac{1}
{2}}.
\end{eqnarray}

In both cases we conclude that the typical range of time for our model to
be healthy is of order $\sqrt{f_1/\Lambda}$ provided that
$f_0$ and $f_1$ are of the same order.

Note that it is known that the theory under consideration admits also de
Sitter type solution\cite{Nojiri:2007uq} with or without the initial large
cosmological constant. The proposed scenario for the effective screening of
cosmological constant works also in this case. It will be considered
elsewhere.

\section{Discussion}

Let us consider a specific example for the values of the model parameters.
Let us assume that the ``bare'' cosmological constant $\Lambda$ in the
action (\ref{nl1}) is very large, say of the order of Planck scale, or of
GUT scale $|\Lambda|\kappa^2 \sim 1-10^{-12}$.
Compared to these values, the cosmological
constant that could explain the accelerated expansion of the present
universe is negligibly small,
$\Lambda_0\kappa^2\sim (10^{-3}\,\mathrm{eV})^4/(10^{18}\,\mathrm{Gev})^4\sim10^{-120}$.
So we may regard the present universe as a flat spacetime at leading order
approximation. In this case, however, we would need our universe to be
stable for
a sufficiently long time, or the ghost-free period should be large enough,
$\Delta t\gtrsim (\Lambda_0)^{-1/2}\sim 10^{60}\kappa$.

For definiteness let us consider the case of $\Lambda>0$.
If we rewrite Eq.~(\ref{ghostsum}) as
\begin{eqnarray}
\frac{1}{f_1}-24<\left(\frac{f_0}{2f_1}+1\right)^2<\frac{1}{f_1}\,,
\end{eqnarray}
we immediately see that this condition is satisfied if $f_1\gg1$
and $f_0=-2f_1$. That is,
\begin{eqnarray}
f(\psi)=f_1\left[(\psi-1)^2-1\right]\,.
\end{eqnarray}
Then from Eqs.~(\ref{psi})-(\ref{ein2}),
the time interval $\Delta t_+$ becomes
\begin{eqnarray}
\Delta
t_+=2\sqrt{\frac{f_1}{\Lambda}}
\left(24-\frac{1}{f_1}\right)^\frac{1}{2}\approx10\sqrt{\frac{f_1}{\Lambda}}
.
\end{eqnarray}
Thus to have $\Delta t\gtrsim (\Lambda_0)^{-1/2}\sim 10^{60}\kappa$
we would need $f_1$ to be extremely large. In this sense, unfortunately
our model cannot be regarded as a complete solution to the cosmological
constant
problem, since we need an unnaturally large number in the model.
Nevertheless, it gives us a hope that for a more sophisticated model
of non-local gravity, it may be possible to have a solution that
screens the bare cosmological constant without encountering the
problem of ghosts.

\begin{acknowledgments}
This work was initiated while all of us were attending
the Long-term Workshop on Gravity and Cosmology
(GC2010: YITP-T-10-01) at Yukawa Institute, Kyoto University.
We would like to thank all the participants of the workshop
for fruitful discussions.
This work was supported in part by
Korea Institute for Advanced Study under the KIAS Scholar program,
by the Grant-in-Aid for the Global COE Program
``The Next Generation of Physics, Spun from Universality and Emergence''
from the Ministry of Education, Culture,
Sports, Science and Technology (MEXT) of Japan,
by JSPS Grant-in-Aid for Scientific Research (A) No.~21244033,
by JSPS Grant-in-Aid for Creative Scientific Research No.~19GS0219,
and by MEC (Spain) project FIS2006-02842.
\end{acknowledgments}

\appendix

\section{Non-local gravity with Lagrange constraint multiplier}

The idea proposed in \cite{Horava:2009uw} for quantum gravity is to
modify the ultraviolet
behavior of the graviton propagator in Lorentz non-invariant way as
$1/\left|\bm{k}\right|^{2z}$, where $\bm{k}$ is the spatial momenta and $z$
could be 2, 3 or larger integers.
They are defined by the scaling properties of space-time coordinates
$\left(\bm{x},t\right)$ as follows,
\be
\label{sym1}
\bm{x}\to b\bm{x}\, ,\quad t\to b^z t\, .
\ee
When $z=3$, the theory seems to be (power counting) UV renormalizable.
Then in order to realize the Lorentz non-invariance, one introduces the
terms breaking the Lorentz invariance explicitly (or more precisely,
breaking
full diffeomorphism invariance) by treating the temporal coordinate and the
spatial
coordinates in a different way.
Such model has the diffeomorphism invariance with respect only to the time
coordinate $t$ and spatial coordinates $\bm{x}$:
\be
\label{sym2}
\delta x^i=\zeta^i(t,\bm{x})\ ,\quad \delta t=f(t)\, .
\ee
Here $\zeta^i(t,\bm{x})$ and $f(t)$ are arbitrary functions.

In \cite{Nojiri:2009th}, Ho\v{r}ava-like gravity model with full
diffeomorphism invariance has been proposed.
When we consider the perturbations from the flat background, which has
Lorentz invariance, the Lorentz invariance of the propagator is
dynamically broken by the
non-standard coupling with a perfect fluid. The obtained propagator behaves
as
$1/{\bm{k}}^{2z}$ with $z=2,3,\cdots$ in the ultraviolet region and the
model could
be perturbatively power counting (super-)renormalizable if $z\geq 3$.
In \cite{Nojiri:2010tv} it was proposed the model of covariant and
power-counting renormalizable field theory of the gravity. The essential
element of the construction is Lagrange multiplier constraint
 proposed in \cite{Lim:2010yk}.

Let us show here that our non-local gravity maybe generalized for the case
of introduction of such Lagrange multiplier constraint in an easy way.


Under the constraint,
\be
\label{LagHL2}
\frac{1}{2} \partial_\mu \phi \partial^\mu \phi
+ U_0 = 0\, ,
\ee
we now define
\bea
\label{nlHrv1}
R^{(2n+1)} &\equiv & R - 2\kappa^2\alpha \left\{
\left(\partial^\mu \phi \partial^\nu \phi \nabla_\mu \nabla_\nu
+ 2 U_0 \nabla^\rho \nabla_\rho \right)^n
\left( \partial^\mu \phi \partial^\nu \phi R_{\mu\nu} + U_0 R
\right)\right\}^2 \, , \nn
R^{(2n+2)} &\equiv & R - 2\kappa^2 \alpha \left\{
\left(\partial^\mu \phi \partial^\nu \phi \nabla_\mu \nabla_\nu
+ 2 U_0 \nabla^\rho \nabla_\rho \right)^n
\left( \partial^\mu \phi \partial^\nu \phi R_{\mu\nu} + U_0 R
\right)\right\}  \nn
&& \times \left\{
\left(\partial^\mu \phi \partial^\nu \phi \nabla_\mu \nabla_\nu
+ 2 U_0 \nabla^\rho \nabla_\rho \right)^{n+1}
\left( \partial^\mu \phi \partial^\nu \phi R_{\mu\nu} + U_0 R
\right)\right\}\, , \nn
\Box^{(n)} &\equiv & \Box + \beta \left(\partial^\mu \phi \partial^\nu \phi
\nabla_\mu \nabla_\nu
+ 2 U_0 \nabla^\rho \nabla_\rho \right)^n \, .
\eea
In the same way as Eq.~(\ref{nl1}), one may define the non-local action,
\be
\label{nlHrv2}
S=\int d^4 x \sqrt{-g}\left\{
\frac{1}{2\kappa^2}\left\{ R^{(m)}\left(1 +
f\left(\left(\Box^{n}\right)^{-1}R^{(m)}\right)\right)
 - 2\Lambda \right\}
 - \lambda \left( \frac{1}{2} \partial_\mu \phi \partial^\mu \phi
+ U_0 \right) + \mathcal{L}_\mathrm{matter} \right\}\, .
\ee
and rewrite the action (\ref{nlHrv2}) in a local way by introducing two
scalar fields $\eta$ and $\xi$:
\bea
\label{nlHrv3}
S&=&\int d^4 x \sqrt{-g}\left[
\frac{1}{2\kappa^2}\left\{R^{(m)} \left(1 + f(\eta)\right)
+ \xi\left(\Box^{(n)} \eta - R^{(m)}\right) -2 \Lambda \right\}
 - \lambda \left( \frac{1}{2} \partial_\mu \phi \partial^\mu \phi
+ U_0 \right) + \mathcal{L}_\mathrm{matter}  \right] \, .
\eea
In (\ref{nlHrv2}) and (\ref{nlHrv3}), $n$ can be either
even or odd integer.
This (power-counting) non-local gravity also admits a flat space solution
 in the presence of cosmological constant. Such a solution has similar
properties as in the model of IV section and may also describe the screening
of cosmological constant.

\end{document}